\documentclass[aps,prl, superscriptaddress, twocolumn, preprintnumbers,amsmath,amssymb,nobalancelastpage,10pt,longbibliography]{revtex4-1}

\usepackage{amsmath}
\usepackage{verbatim}
\usepackage{graphicx}
\usepackage{color}
\usepackage{braket}
\usepackage{yfonts}
\usepackage{soul}

\usepackage{dsfont}

\newcommand{\cop}{\hat{c}}
\newcommand{\cdop}{\hat{c}^\dag}
\newcommand{\Hop}{\hat{H}}
\newcommand{\nop}{\hat{n}}
\newcommand{\llangle}{\left\langle\!\left\langle}
\newcommand{\rrangle}{\right\rangle\!\right\rangle}
\newcommand{\abs}[1]{\lvert {#1} \rvert }
\renewcommand{\Im}{\text{Im}}

\usepackage[colorlinks=true,citecolor=blue,linkcolor=blue,urlcolor=blue]{hyperref}
\usepackage{subfigure}

\usepackage{layouts}

\begin{document}

\title{Self-Trapped Polarons and Topological Defects in a Topological Mott Insulator}
\author{Sergi Juli\`a-Farr\'e}
\email{sergi.julia@icfo.eu}
\affiliation{ICFO - Institut de Ciencies Fotoniques, The Barcelona Institute of Science and Technology, Av. Carl Friedrich Gauss 3, 08860 Castelldefels (Barcelona), Spain}
\author{ Markus M\"uller}
\affiliation{Institute for Quantum Information, RWTH Aachen University, D-52074 Aachen, Germany}
\affiliation{Peter Gr\"unberg Institute, Theoretical Nanoelectronics, Forschungszentrum J\"ulich, D-52428 J\"ulich, Germany}
\author{Maciej Lewenstein}
\affiliation{ICFO - Institut de Ciencies Fotoniques, The Barcelona Institute of Science and Technology, Av. Carl Friedrich Gauss 3, 08860 Castelldefels (Barcelona), Spain}
\affiliation{ICREA, Pg. Llu\'is Companys 23, 08010 Barcelona, Spain}
\author{Alexandre Dauphin}
\email{alexandre.dauphin@icfo.eu}
\affiliation{ICFO - Institut de Ciencies Fotoniques, The Barcelona Institute of Science and Technology, Av. Carl Friedrich Gauss 3, 08860 Castelldefels (Barcelona), Spain}

\begin{abstract}
	
Many-body interactions in topological quantum systems can give rise to new phases of matter, which simultaneously exhibit both rich spatial features and topological properties. In this work, we consider spinless fermions on a checkerboard lattice with nearest and next-to-nearest neighbor interactions. We calculate the phase diagram at half filling, which presents, in particular, an interaction-induced quantum anomalous Hall phase. We study the system at incommensurate fillings using an unrestricted Hartree-Fock ansatz and report a rich zoo of solutions such as self-trapped polarons and domain walls above an interaction-induced topological insulator. We find that, as a consequence of the interplay between the interaction-induced topology and topological defects, these domain walls separate two phases with opposite topological invariants and host topologically protected chiral edge states. Finally, we discuss experimental prospects to observe these novel phenomena in a quantum simulator based on laser-dressed Rydberg atoms in an optical lattice. 
\end{abstract}

\maketitle
{\textit{Introduction}}.\textemdash First proposed by Feynman in the 1980s~\cite{Feynman_1982}, quantum simulators are now a reality. These versatile platforms allow for the simulation of complex quantum many-body systems in a clean and highly controllable environment~\cite{lewenstein2017}. In particular, cold atoms in optical lattices, with the dramatic advances in atomic, molecular, and optical physics, are highly suitable quantum simulators of many-particle or spin systems with controlled interactions~\cite{Bloch_2008}.
There, the study of topological insulators with quantum simulators has become a subject of intense research within the past decade~\cite{Goldman_2014,Cooper_2019}. These exotic materials constitute a new paradigm of quantum matter~\cite{RevModPhys.82.3045}: they are characterized by a global order parameter, an integer called topological invariant, and present topologically protected surface currents. The quantum simulation of such phases typically relies on the generation of artificial gauge fields through Floquet engineering~\cite{Miyake_2013,Aidelsburger_2014,Asteria_2018,Jotzu_2014} or synthetic dimensions~\cite{Celi_2014,Stuhl_2015,Mancini1510}. 

Recent studies focused on the interplay between external gauge fields and interactions~\cite{rachel_2010,Varney2010,PhysRevLett.116.225305,PhysRevB.99.121113}. More strikingly, it has been shown that it is also possible to induce topology directly from interactions, hence giving rise to a spontaneous symmetry-breaking (SSB) topological phase~\cite{Rachel_2018}. For example, interactions of the same order of magnitude between nearest neighbors (NN) and next-to-nearest neighbors (NNN) give rise to the celebrated topological Mott insulator (TMI)~\cite{RaghuTMI2008, Sun2008TRS}, an anomalous quantum Hall (QAH) phase~\cite{Haldane88}, in diverse geometries, such as hexagonal, Lieb, checkerboard, and kagome lattices~\cite{RaghuTMI2008, PhysRevA.86.053618, PhysRevA.93.043611, naturequadraticbands, PhysRevLett.103.046811, PhysRevB.98.125144, PhysRevLett.117.066403, PhysRevLett.103.046811,grushin2011,grushin2013,PhysRevB.82.085310}. To observe these phases, control over the ratio of interaction strengths is crucial, and cold atoms constitute a prime candidate to simulate such phases in experiment~\cite{PhysRevA.86.053618,PhysRevA.93.043611}.
\begin{figure}[t]
\includegraphics[width=\columnwidth]{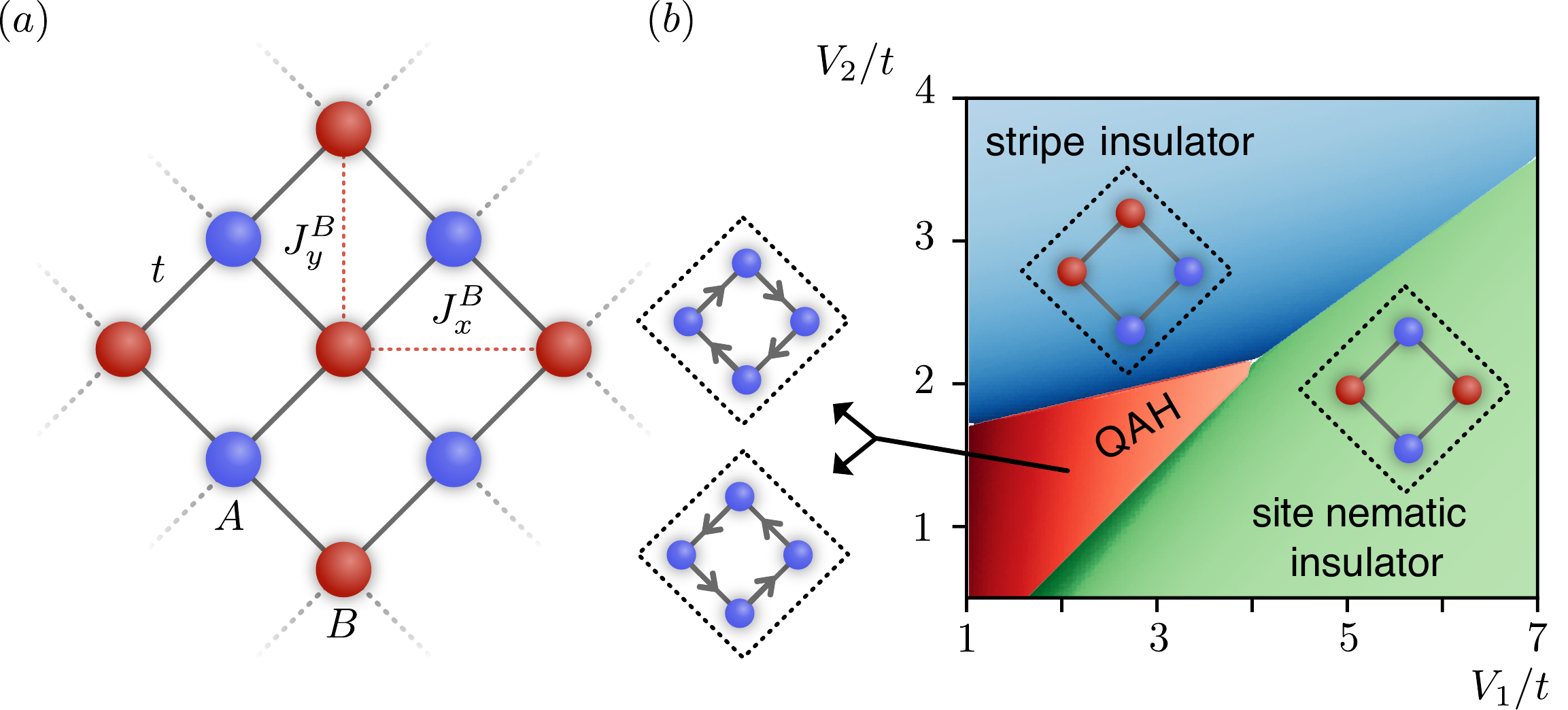}
\caption{(a) The system is described by a Hamiltonian of spinless fermions $\Hop$ on a checkerboard lattice, which has NN and NNN hoppings and interactions. (b) Mean-field phase diagram of the system at half filling in terms of the NN ($V_1$) and NNN ($V_2$) interactions. Each SSB phase is represented with a schematic in the four-sites unit-cell ansatz. For the QAH phase, the two degenerate SSB solutions are explicitly shown.}
\label{fig:intro}
\vspace{-1.9em}
\end{figure}

In this Letter, we explore the exotic nature of self-consistent solutions, such as polarons and topological defects, above the TMI. Remarkably, these solutions appear thanks to the interplay of the global topological order and the SSB local order parameter, and are thus absent in gauge-induced topological insulators. In a SSB lattice material, incommensurate fillings can favor static solutions breaking the translational symmetry \cite{Verges91, Verges92, PhysRevB.83.165118}. These solutions, which can also be created in a dynamical way~\cite{Kibble_1976, Zurek_1985, DelCampo_2014, Polk_2005, Zurek_2005, Keesling_2019}, can take the form of small perturbations, or topological defects in the local order parameter. In the TMI phase, we are interested in studying the impact of these inhomogeneities on the global topology of the system, and vice versa. In order to shed light onto this question, we abandon the usual assumption of spatially homogeneous TMI phases \cite{RaghuTMI2008, PhysRevA.86.053618, PhysRevA.93.043611, naturequadraticbands, PhysRevLett.103.046811, PhysRevB.98.125144, PhysRevLett.117.066403, PhysRevLett.103.046811,grushin2011,grushin2013,PhysRevB.82.085310} and report two type of solutions: (i) self-trapped polarons on top of the topological background and (ii) domain walls between regions from a different sector of the SSB phase. Interestingly, these lead to opposite topological invariants in the same material, and topologically protected chiral edge states at the domain boundaries.  

We start by quantitatively studying the Hartree-Fock (HF) phase diagram of a checkerboard lattice of spinless fermions with NN and NNN interactions at half filling. Such phase diagram is in qualitative agreement with the ones obtained with density matrix renormalization group methods~\cite{PhysRevB.98.125144, naturequadraticbands}. We then study the ground state (GS) around half filling in the QAH phase, with an unrestricted Hartree-Fock (UHF) ansatz. The latter is motivated by previous studies in strongly correlated materials, in which UHF methods describe the main physics of the doped GS, up to corrections in the local quantities~\cite{Poilblanc89, Zaanen89, Machida89, Schulz89, Zheng1155}. At low particle doping, we observe a self-trapped polaron and study its effect on the background topological SSB phase. We also analyze the accuracy of the HF method, and the interaction between two polarons. For higher fillings, we report the appearance of topological defects in the form of domain walls separating the two sectors of the SSB phase at half filling, and inspect the topological chiral edge states on top of them. Finally, we discuss prospects of realizing such phases with cold Rydberg-dressed atoms in optical lattices.

{\textit{Model}}.\textemdash We consider a Hamiltonian of spinless fermions on a checkerboard lattice with periodic boundary conditions, depicted in Fig.~\ref{fig:intro}(a), $\Hop=\Hop_0+\Hop_\text{int}$, where $\Hop_0$ is the quadratic part and $\Hop_{\text{int}}$ contains the interactions. On the one hand, the quadratic part of the fermionic Hamiltonian~\cite{PhysRevLett.103.046811,PhysRevB.98.125144,naturequadraticbands}  reads ($\hbar=1$)
\begin{equation} \label{eq:hamiltonianfree}
\begin{split}
\Hop_{0} = &-t\sum_{\langle ij\rangle}(\cdop_{i,A}\cop_{j,B}+\textrm{H.c.}) \\
&+\sum_{i}\sum_{\substack{\alpha=A,B \\ \nu=x,y}}(J_{\nu}^{\alpha}\cdop_{i,\alpha}\cop_{i+2\nu,\alpha}+\textrm{H.c.})-\mu\sum_{i}\hat{n}_i,
\end{split}
\end{equation}
where $t$ stands for the NN hopping amplitude between sites $A$ and $B$, $J_{\nu}^{\alpha}$ is the NNN hopping amplitude, with $\nu=x, y$  and $\alpha= A, B$, and $\mu$ is the chemical potential that controls the fermionic filling. In the remainder of the Letter, we also set $t=1$, and $J_x^A=J_y^B=+0.5$ and $J_y^A=J_x^B=-0.5$, which generates a $\pi$ flux on each sublattice. Such Hamiltonian has two bands with a quadratic band touching at half filling. On the other hand, the interaction part of the Hamiltonian reads
\begin{equation} \label{eq:hamiltonianinter}
\Hop_\text{int}= V_1\sum_{\langle ij\rangle} \hat{n}_{i,A}\hat{n}_{j, B}+V_2\sum_{\llangle ij\rrangle} \hat{n}_{i}\hat{n}_{j}
\end{equation}
and has NN and NNN density-density repulsive interactions.  Such Hamiltonian has a rich phase diagram in terms of $V_1/t$ and $V_2/t$ already at the mean-field level, as will be discussed below. 

{\textit{Phase diagram at half filling}}.\textemdash We treat the interaction Hamiltonian with a standard HF decoupling, respecting the Wick's theorem, 
\begin{equation} 
\begin{split}
\hat{n}_i\hat{n}_j \simeq \bar{n}_i\nop_j+\bar{n}_j\nop_i-\bar{n}_i\bar{n}_j-\xi_{ij}\cdop_j\cop_i-\xi_{ij}^*\cdop_i\cop_j+\abs{\xi_{ij}}^2,
\end{split}
\end{equation}
with $\xi_{ij}=\langle\cdop_i\cop_j\rangle$ and $\bar{n}_i\equiv\braket{\nop_i}$. The HF values $\xi_{ij}$ and $\bar{n}_i$ are found by determining the self-consistent eigenstates $\lambda$ and energies $E$ of the HF Hamiltonian at zero temperature. We also work with a four-site unit-cell translationally invariant ansatz~[see Fig.~\ref{fig:intro}(b)],  which we will refer to as the restricted Hartree-Fock (RHF) ansatz in the remainder of the Letter~(see Supplemental Material~\cite{SM}). The phase diagram, presented in Fig.~\ref{fig:intro}(b), has three insulating phases, each of them with an order parameter that captures a broken symmetry:  (i) the site nematic insulating phase characterized by $\rho_{n}\equiv |\bar{n}_{A_1}+\bar{n}_{A_2}-\bar{n}_{B_1}-\bar{n}_{B_2}|$, (ii) the stripe insulating phase, with $\rho_s \equiv |\bar{n}_{A_1}-\bar{n}_{A_2}|+|\bar{n}_{B_1}-\bar{n}_{B_2}|$, and (iii) a QAH phase, with time-reversal symmetry breaking (TRSB) due to interaction-induced closed loops of imaginary NN hopping $\xi_\text{QAH}\equiv |\Im ({\xi_{A_1B_1}+\xi_{B_1A_2}+\xi_{A_2B_2}+\xi_{B_2A_1}})|/4$. 
In order to characterize the topology of the QAH phase, we obtain its RHF band structure, which shows two lower filled bands separated from the two upper bands by an energy gap $\Delta_\text{QAH} = 8V_1\xi_\text{QAH}$. We compute the total Chern number $\nu$ of the filled bands and find $\nu=\pm 1$\cite{SM}. The two possible values of $\nu$ account for the twofold ground-state degeneracy in the interaction-induced QAH phase, i.e., the current loops can flow in two opposite directions, and the system reaches one of the two symmetry-breaking sectors through a spontaneous TRSB mechanism [see Fig.~\ref{fig:intro}(b)]. 
 
\begin{figure}[t]
\includegraphics[width=1\columnwidth]{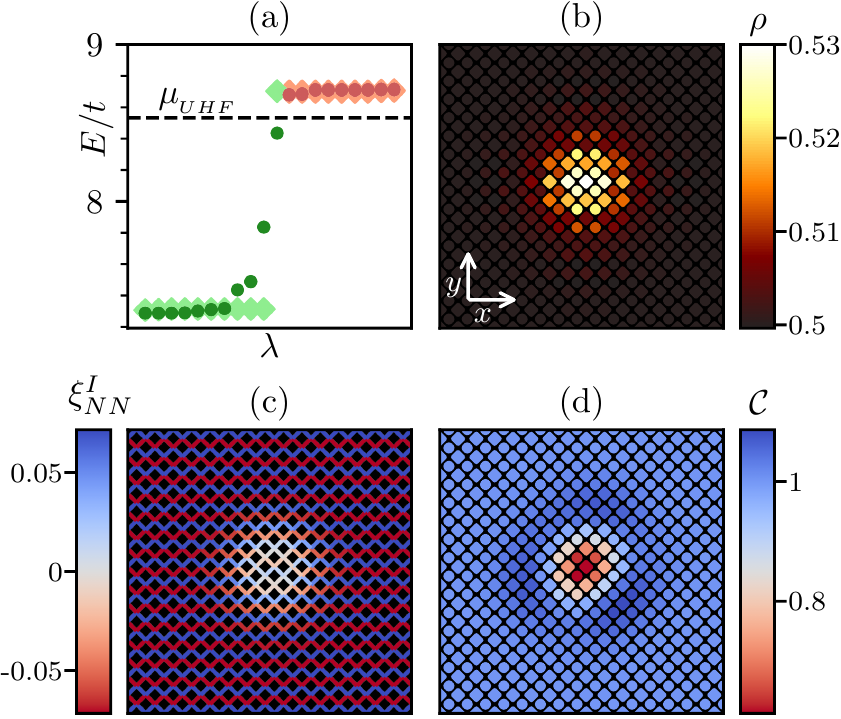}
\caption{Self-trapped polaron for $\delta=1$. (a) Energies of the eigenstates $\lambda$ around the energy gap for the RHF (diamonds) and UHF (dots) ansatz. The dashed line indicates the chemical potential of the UHF solution, and green (red) colors are used for occupied (empty) states. (b)--(d)  Real-space quantities in a region of $16\times16$ unit cells. (b) Density profile. (c) Imaginary NN hopping $\text{Im}\ \xi_{i,i+x\pm y}$. (d) Local Chern number. }
\label{fig:polaron}
\vspace{-2em}
\end{figure} 
{\textit{Self-trapped polaron}}.\textemdash We now focus on the system at low particle doping $\delta$, starting for the case of one extra particle ($\delta=1$). In the noninteracting rigid band picture, the bulk of the system would lose its insulating character, as the particle would occupy the first single-particle state above the gap. In order to analyze the interacting system, hereafter we fix the value of the interactions to $V_1/t=2.5$ and $V_2/t=1.5$. We first study the solution of the RHF ansatz, shown in Fig.~\ref{fig:polaron}(a).  The self-consistent band structure exhibits only a slight deformation of the bands due to interactions and, indeed, we observe the occupation of a single-particle state above the gap. However, there is also the possibility of lowering the energy by creating states inside the gap, which need to be localized in a finite region of the lattice. In order to capture this scenario, we go one step further and work without the requirement of spatial translational invariance of the HF parameters ($\xi_{ij}, \bar{n}_i$), known as the UHF ansatz~\cite{Pratt_1956, Fukutome_1981}, on a $24\times 24$ unit-cell lattice. The results are shown in Fig.~\ref{fig:polaron}(a): the UHF bands are deformed in order to accommodate the extra particle, with the appearance of states inside the insulating gap, and a decrease of the energy $\Delta E_\text{UHF}\equiv \braket{H}_\text{UHF}-\braket{H}_\text{RHF}=-0.12t$. These midgap states are localized in a finite region of the lattice, leading to the density cloud shown in Fig.~\ref{fig:polaron}(b). We denote such state as the self-trapped polaron solution, as the added particle is confined due to its interaction with the topological background. Notice that this notion of a self-trapped polaron differs from the concept of a mobile polaron quasiparticle in a topological background~\cite{PietroImpurityChern}. Indeed, in this work we focus only on static properties, as it is not warranted that a study of the polaron mobility can be made within a coherent quasiparticle picture due to the interaction-induced localization. 

The appearance of self-trapped polarons and other localized excitations has been extensively studied using the UHF method~\cite{Verges91, Verges92, PhysRevB.83.165118} for nontopological phases. We here show that the topological character of the material leads to very interesting physics. On the one hand, Fig.~\ref{fig:polaron}(c) shows that, in the polaron region, there is a large reduction of the order parameter $\xi_\text{QAH}$, accompanied by a change of the SSB sector in the inner region \cite{SM}. Notice that this behavior is similar to the one exhibited by the self-trapped antiferromagnetic polaron observed in the 2D Fermi-Hubbard model~\cite{Verges91}, and can be understood as a collapsed domain wall. On the other hand, and despite the lack of translational invariance, one can characterize the topology of the system, by means of the local Chern number $\mathcal{C}$~\cite{PhysRevB.84.241106, BiancoThesis,PhysRevB.91.085125}.The latter is a real-space quantity that exhibits the same integer quantization as the Chern number $\nu$ within the bulk of the material~\cite{SM}. As shown in Fig.~\ref{fig:polaron}(d), this quantity is not quantized at all inside the polaron; however, it stabilizes to $\mathcal{C}=+1$ further away from the latter.  We emphasize that, as in the half-filling case, the sign of $\mathcal{C}$ depends on the ground-state SSB sector. Notice also that these local fluctuations of $\mathcal{C}$, which are caused by a spontaneous breaking of translational symmetry, are reminiscent of those induced by quenched disorder in a gauge-induced Chern insulator~\cite{ulakar2020kibblezurek}.

{\textit{Configuration interaction analysis}}.\textemdash  We use the configuration interaction method~\cite{Guinea99, PhysRevB.83.165118, PhysRevB.89.155141, SlaterOverlap} to analyze the stability of the polaron localization within the UHF ansatz. That is, for different initial conditions on the UHF parameters, we get degenerate polarons centered at different sites, and it is important to check that such localization is not an artifact of the method. In a nutshell, in the configuration interaction method, one lowers the UHF energy by hybridizing several UHF solutions to restore some of the lattice symmetries spontaneously broken in each UHF solution. In our case, we restore the translational invariance of the polaron solution in smaller checkerboard lattices with up to $2\times 9 \times 9$ sites \cite{SM}. Our analysis yields a polaron band supporting the validity of our UHF treatment: the minimum energy of the band is similar to the UHF energy (with a reduction of $\simeq 0.1 \%$), and this energy corresponds to a plateau of degenerate states in a region of the band of size $|\Delta \mathbf{k}|$. The latter is compatible with a polaron extended over a finite region with radius $\ell_{p}\simeq 1/|\Delta \mathbf{k}|$, as observed in the UHF solution.
\begin{figure}[b]
\centering
\includegraphics[width=
\columnwidth]{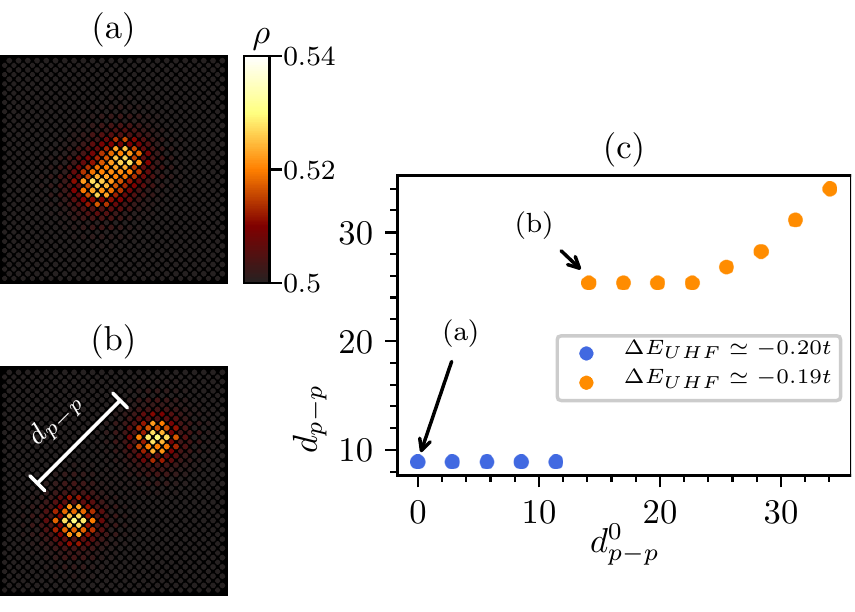}
\caption{Two polarons for $\delta=2$. (a),(b) Density profile of the composite and nonoverlapping solutions.  (c) Final distance between the two polarons, taken as the separation between the two sites with largest density, as a function of the separation in the initial ansatz for the UHF values. Blue (orange) circles correspond to composite (nonoverlapping) solutions.}
\label{fig:2particles}
\vspace{-1.7em}
\end{figure}

{\textit{Two polarons}}.\textemdash For two extra particles ($\delta=2$) we find two types of self-consistent solutions. The lowest energy solution is a composite state of two polarons [see Fig.~\ref{fig:2particles}(a)]. The inner region of this bipolaron, slightly larger than the one corresponding to the single polaron, also exhibits a change in the SSB sector \cite{SM}. The other type of solutions corresponds to two spatially nonoverlapping polarons [see Fig.~\ref{fig:2particles}(b)]. The latter are metastable solutions with a higher energy than the composite state, which indicates that there is an attractive interaction between polarons. However, for large initial separations, individual polarons are likely to be detectable. 
In order to characterize the formation of these metastable solutions, we choose for the initial UHF values those of a spatial superposition of two single polarons, and vary their initial separation. Figure~\ref{fig:2particles}(c) shows the final distance between them $d_{p-p}$ as a function of their initial separation $d^0_{p-p}$. Considering the self-consistent UHF algorithm as some virtual dynamics for the HF parameters, we observe a collapse radius $d^0_{p-p} \simeq 12$ in the initial separation, below which the two polarons interact until the stabilization of the lowest energy composite solution with $d_{p-p}\simeq9$. At larger initial separations, the system stabilizes the nonoverlapping solution, with a forbidden range $d_{p-p} \in (12,25]$, showing that metastable solutions avoid residual overlaps.
\begin{figure}[t]
\centering
\includegraphics[width=
\columnwidth]{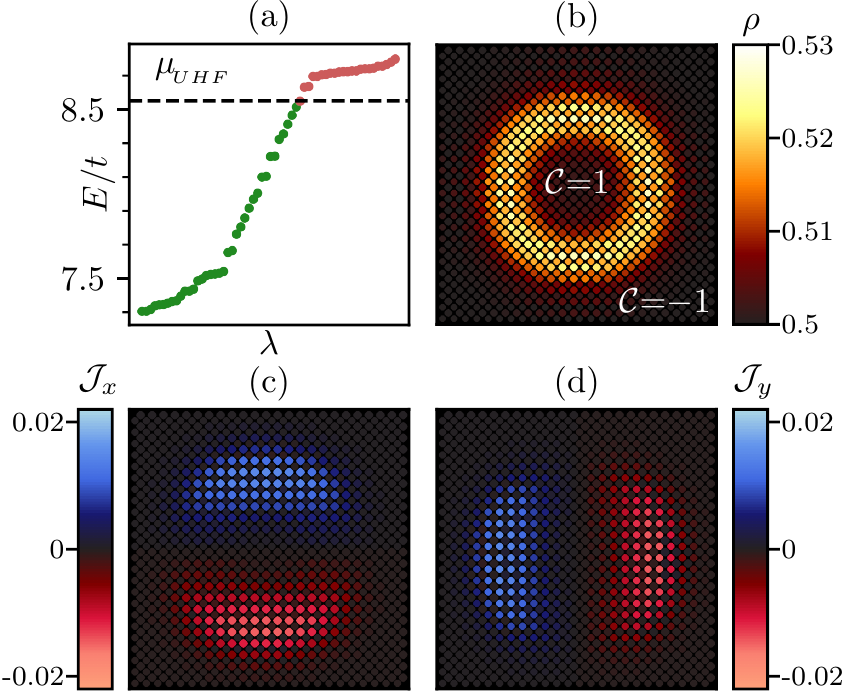}
\caption{Domain wall for $\delta=8$.  This UHF solution lowers the RHF energy by $\Delta E_\text{UHF} \simeq -0.67 t$. (a) Single-particle energies in the gap region. (b) Density profile. Here we indicate the approximately constant value of $\mathcal{C}$ in the inner and outer regions of the ring. (c),(d) GS currents $\mathcal{J}_x$ and $\mathcal{J}_y$.}
\label{fig:8part_ring}
\vspace{-1.7em}
\end{figure}

{\textit{Ring-shaped domain walls}}.\textemdash Even at higher particle doping, we find that the system retains its bulk insulating character: it is energetically favorable to create several midgap localized states, as can be seen in Fig.~\ref{fig:8part_ring}(a) for $\delta=8$.  Taking as a reference the polaron solutions, the attractive interaction between them leads to a GS whose inner region is in the other SSB sector \cite{SM}. For a sufficiently large value of $\delta$, this eventually leads to the formation of a domain wall between two extended half-filling regions that are in the QAH phase, as can be seen in Fig.~\ref{fig:8part_ring}(b). Interestingly, the inversion of the TRSB order parameter across the domain wall leads to the two opposite values of the local Chern number $\mathcal{C}=\pm 1$ depicted in Fig.~\ref{fig:8part_ring}(b). Notice here the role of interactions, leading to a SSB: the coexistence of topologically opposite phases would not occur if the Chern insulator was induced by a homogeneous external gauge field, that would define the global topology sector of the system. Furthermore, the change of the local Chern number $|\Delta \mathcal{C}| = 2$ between the two regions gives rise to topologically protected edge states with a fixed chirality in the ring. This is exactly what we observe in the GS currents $\mathcal{J}_{x(y)}\equiv 2J_{x(y)}^{A/B}\text{Im}\langle \hat{c}^\dagger_{i+x(y)} \hat{c}_i\rangle$ shown in Figs.~\ref{fig:8part_ring}(c) and \ref{fig:8part_ring}(d), which are carried by midgap states. We also verified that the solution with $\mathcal{C}=-1$ in the inner part of the rig and $\mathcal{C}=1$ in the outer part has the same energy and presents edge states with opposite chirality \cite{SM}.

\begin{figure}[t]
\centering
\includegraphics[width=
1.\columnwidth]{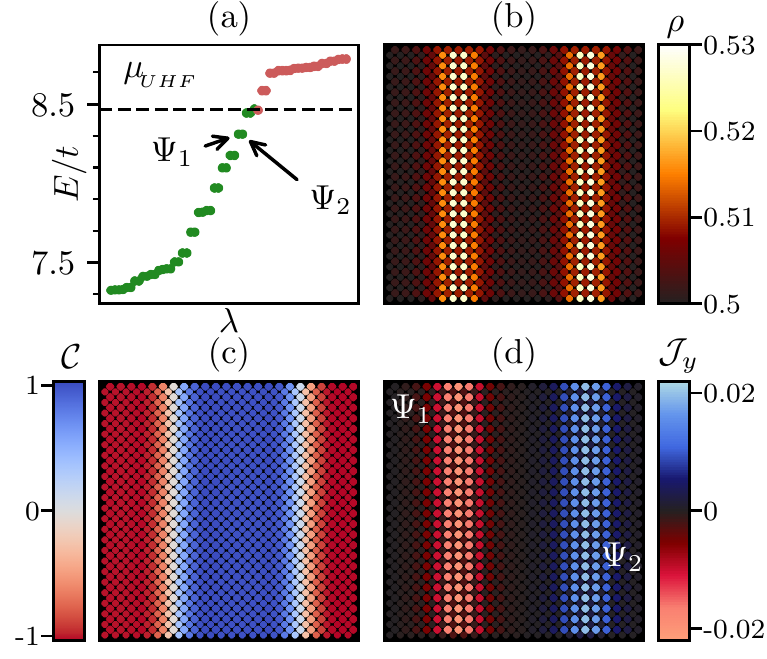}
\caption{Linear domain walls for $\delta=8$. This metastable UHF solution lowers the RHF energy by $\Delta E_\text{UHF}\simeq -0.65t$. (a) Energies in the gap region, with two degenerate edge states $\Psi_1$ and $\Psi_2$. (b) Density profile. (c) Local Chern number. (d)  Current $\mathcal{J}_y$ flowing in opposite directions at each boundary, and carried by midgap states such as $\Psi_1$ and $\Psi_2$.}
\label{fig:8part_linear}
\vspace{-1.7em}
\end{figure}

{\textit{Linear domain walls}}.\textemdash For $\delta=8$, we also find a metastable self-consistent solution, in which the system develops two domain walls (see Fig.~\ref{fig:8part_linear}). The extra density is deposited in these linear structures, as depicted in Fig.~\ref{fig:8part_linear}(b). As in the previous case, the domain walls separate two regions with a reversed TRSB order parameter, leading to two opposite values of the local Chern number, as can be seen in Fig.~\ref{fig:8part_linear}(c). The main difference, however, is that here the change in the local Chern number $|\Delta \mathcal{C}| = 2$ occurs in each of the two disconnected domain walls, leading to pairs of degenerate chiral edge states (one at each domain wall) with opposite chiralities, as shown in Figs.~\ref{fig:8part_linear}(a)-\ref{fig:8part_linear}(d).

{\textit{Quantum simulation with Rydberg atoms in an optical lattice}}.\textemdash Cold atoms in optical lattices provide a prime candidate platform to implement the interacting fermionic Hamiltonian~$\hat{H}$ and observe the interaction-induced polarons and topological defects in a highly controllable experimental setup (for further details, see Ref. \cite{SM}): Laser coupling between different sublattices (cf.~Fig.~\ref{fig:intro}), similar to the approach taken in Ref.~\cite{Goldman_2013}, allows one to control hopping dynamics as described by Eq. \eqref{eq:hamiltonianfree}. Optically trapped and laser-excited Rydberg atoms have become a versatile platform for quantum simulation of many-body physics~\cite{gallagher-book, Browaeys_2020, Bernien_2017}. Weak off-resonant laser dressing of GS atoms with Rydberg-states, as demonstrated, e.g., in Ref.~\cite{Zeiher_2016} allows one to induce effective NN and NNN interactions, with associated energy scale $V_2$ of a similar magnitude as $V_1$~\cite{PhysRevA.86.053618, saffman-rmp-82-2313, Nath-2010,Pupillo-2010, PhysRevA.86.053618, PhysRevA.93.043611}, and both compatible with the kinetic energy scales of Eq.\eqref{eq:hamiltonianfree}. Detrimental incoherent processes, such as dephasing and radiative decay from metastable Rydberg states, are small for suitably chosen laser and atomic parameters~\cite{Balewski_2014} and occur at timescales larger than those associated to the effective couplings of Hamiltonian \eqref{eq:hamiltonianinter} governing the coherent system dynamics. For $t\sim 1$ kHz, temperatures $T\lesssim 10$ nK are below the critical temperature of the QAH phase and the typical decrease in energy due to the appearance of spatial structures inside the bulk gap. In addition, techniques specifically adapted to the detection of the Chern number, such as time of flight~\cite{PhysRevLett.107.235301, Goldman_2013}, transport~\cite{dauphin2013,Aidelsburger_2014}, or interband transition~\cite{Tran2017,Asteria_2018} measurements or the detection of edge states on the domain walls through real-space-density imaging~\cite{Goldman_2013,Goldman_2016}, could be generalized to resolve the predicted interaction-induced polarons and topological defects.

\begin{acknowledgments}
We acknowledge support from ERC AdG NOQIA, Spanish Ministry of Economy and Competitiveness (“Severo Ochoa” program for Centres of Excellence in R\&D (CEX2019-000910-S), Plan National FISICATEAMO and FIDEUA PID2019-106901GB-I00/10.13039/501100011033, FPI), Fundaci\'o Privada Cellex, Fundaci\'o Mir-Puig, and from Generalitat de Catalunya (AGAUR Grant No. 2017 SGR 1341, CERCA program, QuantumCAT\_U16-011424, co-funded by ERDF Operational Program of Catalonia 2014-2020), MINECO-EU QUANTERA MAQS (funded by State Research Agency (AEI) PCI2019-111828-2/10.13039/501100011033), EU Horizon 2020 FET-OPEN OPTOLogic (Grant No 899794), and the National Science Centre, Poland-Symfonia Grant No. 2016/20/W/ST4/00314. A. D. acknowledges the financial support from a fellowship granted by la Caixa Foundation (ID 100010434, fellowship code LCF/BQ/PR20/11770012). M. M. acknowledges support by the ERC StG QNets 804247.
\end{acknowledgments}

\bibliographystyle{apsrev4-1}
\end{document}


\title{Supplemental Materials: Self-trapped polarons and topological defects in a topological Mott insulator}
\author{Sergi Juli\`a-Farr\'e}
\affiliation{ICFO - Institut de Ciencies Fotoniques, The Barcelona Institute of Science and Technology, Av. Carl Friedrich Gauss 3, 08860 Castelldefels (Barcelona), Spain}
\author{ Markus M\"uller}
\affiliation{Institute for Quantum Information, RWTH Aachen University, D-52074 Aachen, Germany}
\affiliation{Peter Gr\"unberg Institute, Theoretical Nanoelectronics, Forschungszentrum J\"ulich, D-52428 J\"ulich, Germany}
\author{Maciej Lewenstein}
\affiliation{ICFO - Institut de Ciencies Fotoniques, The Barcelona Institute of Science and Technology, Av. Carl Friedrich Gauss 3, 08860 Castelldefels (Barcelona), Spain}
\affiliation{ICREA, Pg. Llu\'is Companys 23, 08010 Barcelona, Spain}
\author{Alexandre Dauphin}
\affiliation{ICFO - Institut de Ciencies Fotoniques, The Barcelona Institute of Science and Technology, Av. Carl Friedrich Gauss 3, 08860 Castelldefels (Barcelona), Spain}

\begin{abstract}
	
\end{abstract}

\maketitle
\makeatletter
\renewcommand{\theequation}{S\arabic{equation}}
\renewcommand{\thesection}{S\Roman{section}}
\renewcommand{\thefigure}{S\arabic{figure}}
\renewcommand{\bibnumfmt}[1]{[S#1]}
\renewcommand{\citenumfont}[1]{S#1}

\section{Band structure of the bare Hamiltonian}\label{app:bare}

In order to take advantage of the quadratic and translational invariance of the free  Hamiltonian $\hat{H}_0$  (see Eq.~(1) of the main text), we write it in terms of the $k$-space fermionic operators, defined as
\begin{equation}
c_{k,\alpha} = \frac{1}{\sqrt{L}}\sum_j e^{\im j_\alpha k}c_{j,\alpha},\hspace{8mm} \alpha = {A,\ B}.
\end{equation}
Here the total number of two-site unit cells is given by $L=L_x\times L_y$, and we have $\mathbf{k}=(k_x, k_y)$ with $k_x \in [0,\pi)$, $\Delta k_x = \pi/L_x$, and similarly for $k_y$.  Importantly, $j$ is a tuple of indices $j\equiv(j_x, j_y)$ specifying the position of each unit cell in the 2D lattice. Moreover, notice that $j$ refers only to the unit cell, while the index $j_\alpha$ appearing in the exponent refers to the physical position of the fermionic operator. Therefore, for A sites, $j$ and $j_A$ coincide, whereas for B sites $j_B=j+(1,1)$. In $k$-space, $H_0$  reads
\begin{equation}\label{eq:hamiltonianfreek}
\Hop_0 = \sum_k \boldsymbol{\Psop}_k^\dagger H_0^k\boldsymbol{\Psop}_k,
\end{equation}
with the definitions 
\begin{equation}\label{eq:definitonskspace}
\begin{split}
\boldsymbol{\Psop}_k =& (\cop_{k,A},\ \cop_{k,B})^T,\ H_{0,01}^k =H_{0,10}^k= -4t\cos{k_x}\cos{k_y},\\
H_{0,00}^k=&-2\left[J_x^A\cos(2k_x)+J_y^A\cos(2k_y)\right],\
H_{0,11}^k = -2\left[J_x^B\cos(2k_x)+J_y^B\cos(2k_y)\right].
\end{split}
\end{equation}
Now, Eq.~\eqref{eq:hamiltonianfreek} can be brought into diagonal form 
\begin{equation}\label{eq:hamiltonianfreekdiag}
\Hop_0 = \sum_k \boldsymbol{\Phop}_k^\dagger H_D^k\boldsymbol{\Phop}_k,
\end{equation}
where 
\begin{equation}\label{eq:definitonskbogo}
\boldsymbol{\Phop}_k = (\Phop_{k,0},\ \Phop_{k,1})^T,\hspace{5mm} H_D^k=\begin{pmatrix}
	  E_{k,-} &  0\\
	    0& E_{k,+}
	\end{pmatrix}.
\end{equation}
 Notice that for each $k$ we have two energies corresponding to the two bands of the lattice. Here we give the analytical form of these energy eigenvalues
 \begin{equation}
 \begin{split}
 &E_{k,\pm}=-\left[T_x\cos(2k_x)+T_y\cos(2k_y)\right]\pm \sqrt{\left[\Delta_x\cos(2k_x)+\Delta_y\cos(2k_y)\right]^2+16t^2\cos^2k_x\cos^2k_y},
 \end{split}
 \end{equation}
where we have defined $T_\mu (\Delta_\mu) = J^A_\mu +(-)\ J^B_\mu$. With the choice of parameters used in the main text, we have $t=1$, $T_\mu=0$ and $\Delta_x=1=-\Delta_y$, leading to 
\begin{equation}
 \begin{split}
 &E_{k,\pm}=\pm \sqrt{\left[\cos(2k_x)-\cos(2k_y)\right]^2+16\cos^2k_x\cos^2k_y}.
 \end{split}
 \end{equation}

\section{Hartree-Fock band structure of the interacting Hamiltonian (4-sites unit cell ansatz)}\label{app:interaction_4sites}

\begin{figure}[t]
\centering
\includegraphics{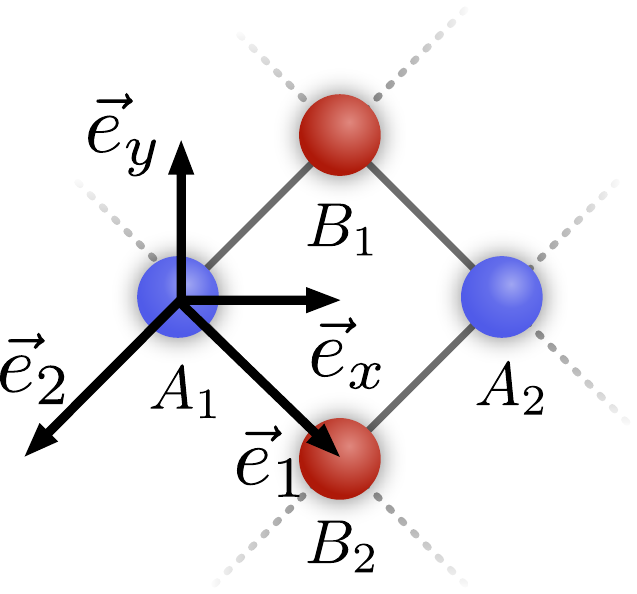}
\caption{Schematic of the 4 unit cell ansatz in the checkerboard lattice.}
\label{fig:unitcell}
\end{figure}

We extend $H_0$ by including density interactions between different sites as follows
\begin{equation}\label{eq:hamiltoniansup}
\Hop=\Hop_0 + V_1\sum_{\langle ij\rangle} \hat{n}_{i,A}\hat{n}_{j, B}+V_2\sum_{\llangle ij\rrangle} \hat{n}_{i}\hat{n}_{j}.
\end{equation}
We take a Hartree-Fock (HF) approach to solve it by means of a self-consistent iterative algorithm at a given filling and zero temperature. We use the following decomposition that respects Wick's Theorem
\begin{equation}
\begin{split}
\hat{n}_i\hat{n}_j \simeq &-\xi_{ij}\cdop_j\cop_i-\xi_{ij}^*\cdop_i\cop_j+\abs{\xi_{ij}}^2+\mean{\nop_i}\nop_j+\mean{\nop_j}\nop_i-\mean{\nop_i}\mean{\nop_j}.
\end{split}
\end{equation}
with $\xi_{ij}=\langle\cdop_i\cop_j\rangle$. We now transform the interaction terms to momentum space, assuming a 4-sites unit-cell ansatz (see Fig.~\ref{fig:unitcell}). In order to allow for loops of imaginary hopping between NNs, we furthermore impose $\langle\cdop_{i,{B_1}}\cop_{i,{A_1}}\rangle=\langle\cdop_{i,{B_1}}\cop_{i-2e_2,{A_1}}\rangle$, $\langle\cdop_{i,{B_2}}\cop_{i,{A_1}}\rangle=\langle\cdop_{i-2e_1,{B_2}}\cop_{i,{A_1}}\rangle$, $\langle\cdop_{i,{A_2}}\cop_{i,{B_1}}\rangle=\langle\cdop_{i-2e_1,{A_2}}\cop_{i,{B_1}}\rangle$, and $\langle\cdop_{i,{A_2}}\cop_{i,{B_2}}\rangle=\langle\cdop_{i-2e_2,{A_2}}\cop_{i,{B_2}}\rangle$. We also restrict the mean field to real $\xi_{NNN}$. We get to the following form for the Hamiltonian \eqref{eq:hamiltoniansup}
\begin{equation}\label{eq:hamiltonianinterk}
\Hop = E_0+\sum_k \boldsymbol{\Psop}_k\dagg H_{V}^k\boldsymbol{\Psop}_k,
\end{equation}
with
\begin{equation}\label{eq:hamiltonianinterkdiag}
\begin{split}
\boldsymbol{\Psop}_k =& (\cop_{k,{A_1}},\ \cop_{k,{B_1}}, \cop_{k,{B_2}}, \cop_{k,{A_2}})^T,\\
H_{V,00}^k=&2V_1(\bar{n}_{B_1}+\bar{n}_{B_2})+4V_2\bar{n}_{A_2},\hspace{3mm} H_{V,11}^k=2V_1(\bar{n}_{A_1}+\bar{n}_{A_2})+4V_2\bar{n}_{B_2},\\
H_{V,22}^k=&2V_1(\bar{n}_{A_1}+\bar{n}_{A_2})+4V_2\bar{n}_{B_1},\hspace{3mm} H_{V,33}^k=2V_1(\bar{n}_{B_1}+\bar{n}_{B_2})+4V_2\bar{n}_{A_1},\\
H_{V,01}^k=&-2(t+V_1\xi^R_{{A_1}{B_1}})\cos(k_2)-2iV_1\xi^I_{{A_1}{B_1}}\cos(k_2),\\
H_{V,02}^k=&-2(t+V_1\xi^R_{{A_1}{B_2}})\cos(k_1)-2iV_1\xi^I_{{A_1}{B_2}}\cos(k_1),\\
H_{V,03}^k=&2(J_x^{A}-V_2\xi^x_{{A_1}{A_2}})\cos(k_1+k_2)+2(J_y^{A}-V_2\xi^y_{{A_1}{A_2}})\cos(k_1-k_2),\\
H_{V,12}^k=&2(J_x^{B}-V_2\xi^x_{{B_1}{B_2}})\cos(k_1+k_2)+2(J_y^{B}-V_2\xi^y_{{B_1}{B_2}})\cos(k_1-k_2),\\
H_{V,13}^k=&-2(t+V_1\xi^R_{{B_1}{A_2}})\cos(k_1)-2iV_1\xi^I_{{A_1}{B_1}}\cos(k_1),\\
H_{V,23}^k=&-2(t+V_1\xi^R_{{B_2}{A_2}})\cos(k_2)-2iV_1\xi^I_{{B_2}{A_2}}\cos(k_2),\\
\frac{E_0}{L_{4}}=&2V_1\left[|\xi_{{A_1}{B_1}}|^2+|\xi_{{B_2}{A_2}}|^2+|\xi_{{B_1}{A_2}}|^2+|\xi_{{A_1}{B_2}}|^2-(\bar{n}_{A_1}+\bar{n}_{A_2})(\bar{n}_{B_1}+\bar{n}_{B_2})\right]+\\
&+2V_2\left[(\xi^x_{{A_1}{A_2}})^2+(\xi^y_{{A_1}{A_2}})^2+(\xi^x_{{B_1}{B_2}})^2+(\xi^y_{{B_1}{B_2}})^2-2(\bar{n}_{A_1}\bar{n}_{A_2}+\bar{n}_{B_1}\bar{n}_{B_2})\right],\\
k_1, k_2 \in &[0,\ \pi).
\end{split}
\end{equation}
Notice that, in this $4-$sites unit-cell ansatz, the reciprocal space components $(k_1, k_2)$  are rotated by $45^{\circ}$ with respect to the $(k_x,k_y)$ components of the previous Sec. \ref{app:bare}.

\section{Hartree-Fock band structure of the interacting Hamiltonian (2-site unit cell ansatz)}\label{app:interaction}

As the quantum anomalous Hall phase can be captured in a simplified ansatz of a 2-site unit cell (sites A and B in Fig. 1a of the main text), here we show the quasi-analytical band structure that can be derived from it. We also start from the interacting Hamiltonian
\begin{equation}\label{eq:hamiltonianinter2}
\Hop=\Hop_0 + V_1\sum_{\langle ij\rangle} \hat{n}_{i,A}\hat{n}_{j, B}+V_2\sum_{\llangle ij\rrangle} \hat{n}_{i}\hat{n}_{j}.
\end{equation}
As in the previous case, we take a Hartree-Fock (HF) approach. We now transform the interaction terms to Fourier space, assuming $\langle\cdop_{i,B}\cop_{j,A}\rangle=|\xi_{BA}|e^{\im \theta}$. The phase $\theta$ fulfills the constraint
\begin{equation}
\begin{split}
\langle\cdop_{i,B}\cop_{i,A}\rangle =&\ \langle\cdop_{i,B}\cop_{i+2\vec{e}_x,A}\rangle^*=\langle\cdop_{i,B}\cop_{i-2\vec{e}_y,A}\rangle^*=\langle\cdop_{i,B}\cop_{i+2\vec{e}_x-2\vec{e}_y,A}\rangle.
\end{split}
\end{equation}
We also set to real $\xi_{Ax/y}$ and $\xi_{Bx/y}$. We get to the following form for the Hamiltonian (\ref{eq:hamiltonianinter2}) 
\begin{equation}\label{eq:hamiltonianinterk2}
\Hop = E_0+\sum_k \boldsymbol{\Psop}_k\dagg H_{V}^k\boldsymbol{\Psop}_k,
\end{equation}
with
\begin{equation}
\begin{split}
\boldsymbol{\Psop}_k =&\ (\cop_{k,A},\ \cop_{k,B})^T,\\
H_{V,00}^k=&-2\left[(J_x^A+V_2\xi_{Ax})\cos(2k_x)+(J_y^A+V_2\xi_{Ay})\cos(2k_y)\right]+4(V_1\bar{n}_B+V_2\bar{n}_A),\\
H_{V,11}^k =&-2\left[(J_x^B+V_2\xi_{Bx})\cos(2k_x)+(J_y^B+V_2\xi_{By})\cos(2k_y)\right]+4(V_1\bar{n}_A+V_2\bar{n}_B),\\
H_{V,01}^k =& -4(t+V_1\xi_{BA}^R)\cos{k_x}\cos{k_y}+\im 4V_1\xi_{BA}^I\sin{k_x}\sin{k_y},=(H_{V,10}^k)^*\\
\frac{E_0}{L} =&\ 4V_1 \left(\abs{\xi_{BA}}^2-\bar{n}_A\bar{n}_B\right)+V_2(\xi_{Ax}^2+\xi_{Ay}^2+\xi_{Bx}^2+\xi_{By}^2-2\bar{n}_A^2-2\bar{n}_B^2),\\
k_x, k_y \in &\ [0,\ \pi).
\end{split}
\end{equation}
To analytically diagonalize the Hamiltonian (\ref{eq:hamiltonianinterk2}), we write it in terms of Pauli matrices as 
\begin{equation}\label{eq:hamiltonian_pauli}
\begin{split}
&\boldsymbol{\Psop}_k\dagg\left(\alpha_k \id + \beta_k \sigma_z +\gamma_k\sigma_x + \lambda_k\sigma_y\right)\boldsymbol{\Psop}_k\equiv \boldsymbol{\Psop}_k\dagg\left(\alpha_k \id + \vec{n}_k\cdot \vec{\sigma}\right)\boldsymbol{\Psop}_k,\\
&\vec{n}_k/\abs{\vec{n}_k}= (\sin\theta \cos\phi,\ \sin\theta \sin\phi, \cos\theta)^T,\ (\sin\theta\ge 0)
\end{split}
\end{equation}
\begin{equation}
\begin{split}
\alpha_k=&-\left[T_x\cos(2k_x)+T_y\cos(2k_y)\right]+2(V_1+V_2)n,\hspace{3mm}\beta_k=-\left[\Delta_x\cos(2k_x)+\Delta_y\cos(2k_y)\right]-2(V_1-V_2)\rho,\\
\gamma_k=&-4(t+V_1\xi_{BA}^R)\cos(k_x)\cos(k_y),\hspace{3mm}\lambda_k=-4V_1\xi_{BA}^I\sin(k_x)\sin(k_y),\\
T_\mu =&\ J_{\mu}^A+J_{\mu}^B+V_2(\xi_{A\mu}+\xi_{B\mu}),\hspace{3mm}\Delta_\mu =\ J_{\mu}^A-J_{\mu}^B+V_2(\xi_{A\mu}-\xi_{B\mu}),\hspace{3mm}n =\ \bar{n}_A+\bar{n}_B,\hspace{3mm}\rho =\ \bar{n}_A-\bar{n}_B.
\end{split}
\end{equation}
We have the following energies and ground-state eigenvector
\begin{equation}
\begin{split} 
E_{k,\pm} =\ \alpha_k \pm \abs{\vec{n}_k},\hspace{5mm}
\vec{v}_{k,-} = \left(e^{-\im\phi_k/2}\sin{\theta_k/2},\ -e^{\im\phi_k/2}\cos{\theta_k/2}\right)^T.
\end{split}
\end{equation}
Now we can compute the expectation values in the ground state (we use the notation $\vec{v}\equiv \vec{v}_{k,-}$)
\begin{equation}
\begin{split}
\rho =& \frac{1}{L}\sum_k \abs{\vec{v}_{00}}^2-\abs{\vec{v}_{11}}^2=\frac{-1}{L}\sum_k \cos(\theta_k)
=\frac{-1}{L}\sum_k \beta_k/\abs{\vec{n}_k},\\
\xi_{BA}^R =& \frac{-1}{2L}\sum_k\cos(k_x)\cos(k_y)\sin(\theta_k)\cos(\phi_k)
=\frac{-1}{2L}\sum_k \cos(k_x)\cos(k_y)\gamma_k/\abs{\vec{n}_k},\\
\xi_{BA}^I=& \frac{1}{2L}\sum_k\cos(k_x)\cos(k_y)\sin(\theta_k)\sin(\phi_k)
=\frac{1}{2L}\sum_k \cos(k_x)\cos(k_y)\lambda_k/\abs{\vec{n}_k},\\
\xi_{Ax(y)}=&\frac{1}{L}\sum_k \cos(2k_{x(y)})\abs{\vec{v}_{00}}^2,\hspace{3mm} \xi_{Bx(y)}=\frac{1}{L}\sum_k \cos(2k_{x(y)})\abs{\vec{v}_{11}}^2,
\end{split}
\end{equation}
and the value of the free energy
\begin{equation}
\begin{split}
F/L =4V_1\left[(\xi_{BA}^R)^2+(\xi_{BA}^I)^2+\frac{1}{4}(\rho^2-1)\right]+V_2\left[\xi_{Ax}^2+\xi_{Ay}^2+\xi_{Bx}^2+\xi_{By}^2-(1+\rho^2)\right]+\frac{1}{L}\sum_k (\alpha_k - \abs{\vec{n}_k}),
\end{split}
\end{equation}
which can be brought into the equivalent form 
\begin{equation}
\begin{split}
\tilde{F}/L = 2V_1\left[(\xi_{BA}^R)^2+(\xi_{BA}^I)^2+\rho^2/4\right]+V_2\left[\xi_{Ax}^2+\xi_{Ay}^2+\xi_{Bx}^2+\xi_{By}^2-\rho^2\right]+\frac{1}{L}\sum_k (\alpha_k - \abs{\vec{n}_k}).
\end{split}
\end{equation}

\section{Chern number formula of the lower band in the 2-sites ansatz}
We can write the Chern number of the checkerboard lattice as

\begin{equation} \label{eq:cherntwoband}
\nu = \frac{\pi}{2L}\sum_{k} \vec{n}_k \cdot \left(\frac{\partial \vec{n}_k}{\partial k_x}\times \frac{\partial \vec{n}_k}{\partial k_y}\right),
\end{equation}
where $\vec{n}_k$ is given in Eq.~\eqref{eq:hamiltonian_pauli}. It is useful to define $\vec{v}_k = (\gamma_k,\ \lambda_k,\ \beta_k)$, such that $\vec{n}_k = \vec{v}_k/\abs{\vec{v}_k}$. Then 

\begin{equation}  
\frac{\partial\vec{n}_k}{\partial k_\mu} = \frac{1}{\abs{\vec{v}_k}}\frac{\partial\vec{v}_k}{\partial k_\mu}-\frac{1}{\abs{\vec{v}_k}^3}\left(\vec{v}_k\cdot\frac{\partial\vec{v}_k}{\partial k_\mu}\right)\vec{v}_k.
\end{equation}

The expressions are the following

\begin{equation}  
\begin{split}
\frac{\partial\vec{v}_k}{\partial k_x}=&\left[4(t+V_1\xi_{BA}^R)\sin(k_x)\cos(k_y),
-4V_1\xi_{BA}^I\cos(k_x)\sin(k_y),2\Delta_x\sin(2k_x)\right],\\
\frac{\partial\vec{v}_k}{\partial k_y}=&\left[4(t+V_1\xi_{BA}^R)\sin(k_y)\cos(k_x),-4V_1\xi_{BA}^I\cos(k_y)\sin(k_x),-2\Delta_x\sin(2k_y)\right].
\end{split}
\end{equation}

Notice that to apply this formula we first have to determine the Hartree-Fock parameters $\xi$'s and $\rho$.

\section{Calculation of the local Chern number}\label{app:chern_loc}
We follow the procedure introduced in \cite{BiancoThesis} and used in \cite{PhysRevB.91.085125} to compute the real space Chern number for a non-homogeneous system. The value of $C$ is quantized to the same value as the Chern number, and a non-zero value indicates the presence of topologically protected chiral states at the edges of the bulk in which $C$ is computed. We first define the projectors onto the occupied and unoccupied single-particle states of the unrestricted Hartree-Fock Hamiltonian
\begin{equation} 
P = \sum_{i\in \text{occ}} \ket{\Phi_i}\bra{\Phi_i}, \hspace{5mm} \hat{Q} = \id -\hat{P}.
\end{equation}
We also define the operators 
\begin{equation}
\begin{split}
\hat{\boldsymbol{r}}_P = \hat{P}\hat{\boldsymbol{r}}\hat{Q}=(\hat{x}_P,\hat{y}_P),\hspace{5mm} 
\hat{\boldsymbol{r}}_Q = \hat{Q}\hat{\boldsymbol{r}}\hat{P}=(\hat{x}_Q,\hat{y}_Q).
\end{split}
\end{equation}
These operators are quasi-local, in the sense that $\bra{i}\hat{r}_{P,Q}\ket{j}\sim \exp{(-\kappa_{P,Q}\norm{i-j})}$. Now, we define the local Chern marker as
\begin{equation}\label{eq:localchern}
\textswab{C}(i)=-4\pi\text{Im}\left[\sum_{j}\bra{i}\hat{x}_Q\ket{j}\bra{j}\hat{y}_P\ket{i}\right]
\end{equation}
with 
\begin{equation}
\bra{i}\hat{x}_Q\ket{j} = \sum_l Q(i,l)x_lP(l,j),\ \bra{j}\hat{y}_P\ket{i} = \sum_l P(j,l)y_lQ(l,i).
\end{equation}
Finally, to compute the Chern number $C$, one has to perform the spatial average of the local Chern marker (\ref{eq:localchern})
\begin{equation}
C = \frac{1}{A_{\text{disk}}}\sum_{i\in \text{disk}} \textswab{C}(i),
\end{equation}
where the disk is the set of points inside a circle of a certain radius, considered to lie inside the bulk of the lattice, where edge effects do not play a role, and $A_{\text{disk}}$ is the area of this disk. In this work, we always set the radius to $r=a/\sqrt{2}$, where $a$ is the distance between NN.

\section{Behavior of the TRSB local order parameter}\label{app:SSB}
\begin{figure}[h!]
\centering
\includegraphics[width=
\columnwidth]{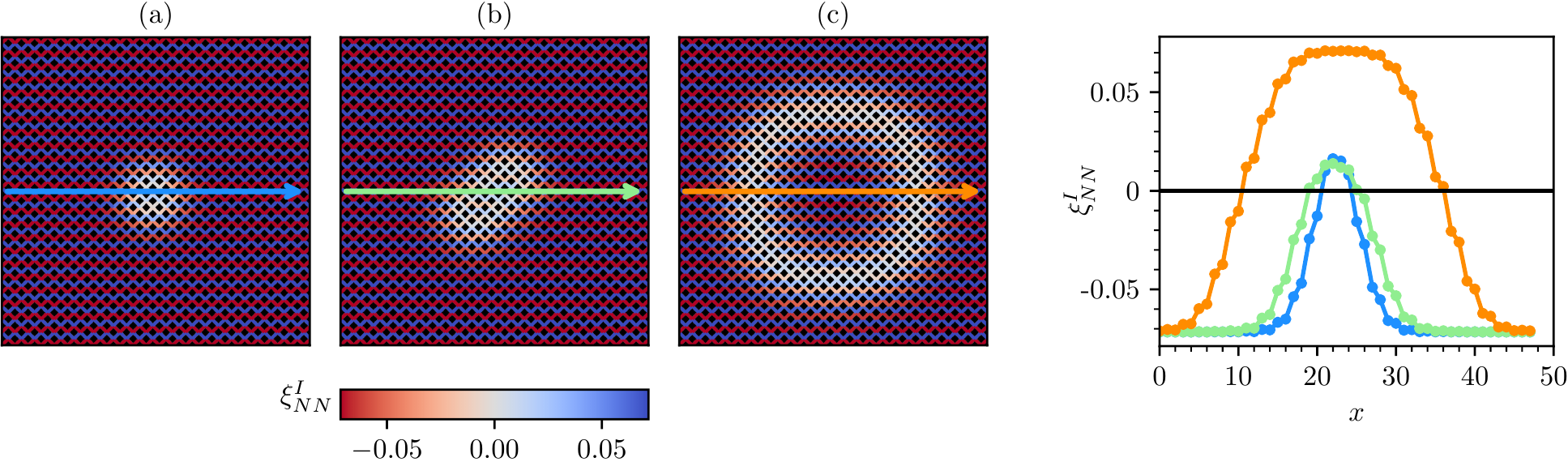}
\caption{Local order parameter of the spontaneous time-reversal symmetry breaking $\xi_{NN}^I$ for different unrestricted Hartree-Fock solutions. (a) Polaron solution obtained for one extra particle ($\delta=1$). (b) Bipolaron solution for $\delta=2$. (c) Ring domain wall solution for $\delta=8$. (d) Plot of $\xi_{NN}^I$ along the horizontal cut depicted in (a)-(c). This quantity reveals the inversion of the SSB sector inside the inhomogeneities.}
\label{fig:SSB}
\end{figure}

Here we analyze the behavior of the imaginary part of the hopping between nearest neighbors $\xi_{NN}^I$ (i.e., the local order parameter of the spontaneous time-reversal symmetry breaking) in the different inhomogeneous solutions discussed in the main text. Figures~\ref{fig:SSB}(a)-(c) show its value in the lattice for the polaron, bipolaron, and ring solutions. We observe the non-zero value of this quantity in the bulk, with different signs indicating the closed current loops characteristic of the quantum Anomalous Hall phase. We also observe a large reduction of the absolute value of $\xi_{NN}^I$ in the polaron, bipolaron and ring regions, coinciding with the region where there is a deviation of the half-filling occupation. To analyze the evolution of $\xi_{NN}^I$ trough the inhomogeneities, in Fig.~\ref{fig:SSB}(d) we plot $\xi_{NN}^I$ in a zig-zag line connecting NN, indicated in Figs.~\ref{fig:SSB}(a)-(c). Figure~\ref{fig:SSB}(d) then shows the transition from the polaron solution to the ring solution in terms of the SSB:
inside the polaron there is change of the SSB sector due to the change in the sign of $\xi_{NN}^I$. However the inner region is too small to develop a well-defined domain of the SSB phase. In the bipolaron case the behavior is similar, with a slight widening of the curve. In the ring solution, however, we observe a full inversion of $\xi_{NN}^I$ in the inner region, leading to two domains belonging to the different sectors of the SSB phase.
 
\section{Spontaneous symmetry breaking sectors of the ring solution}\label{app:ring}
\begin{figure}[h!]
\centering
\includegraphics[width=
0.9\columnwidth]{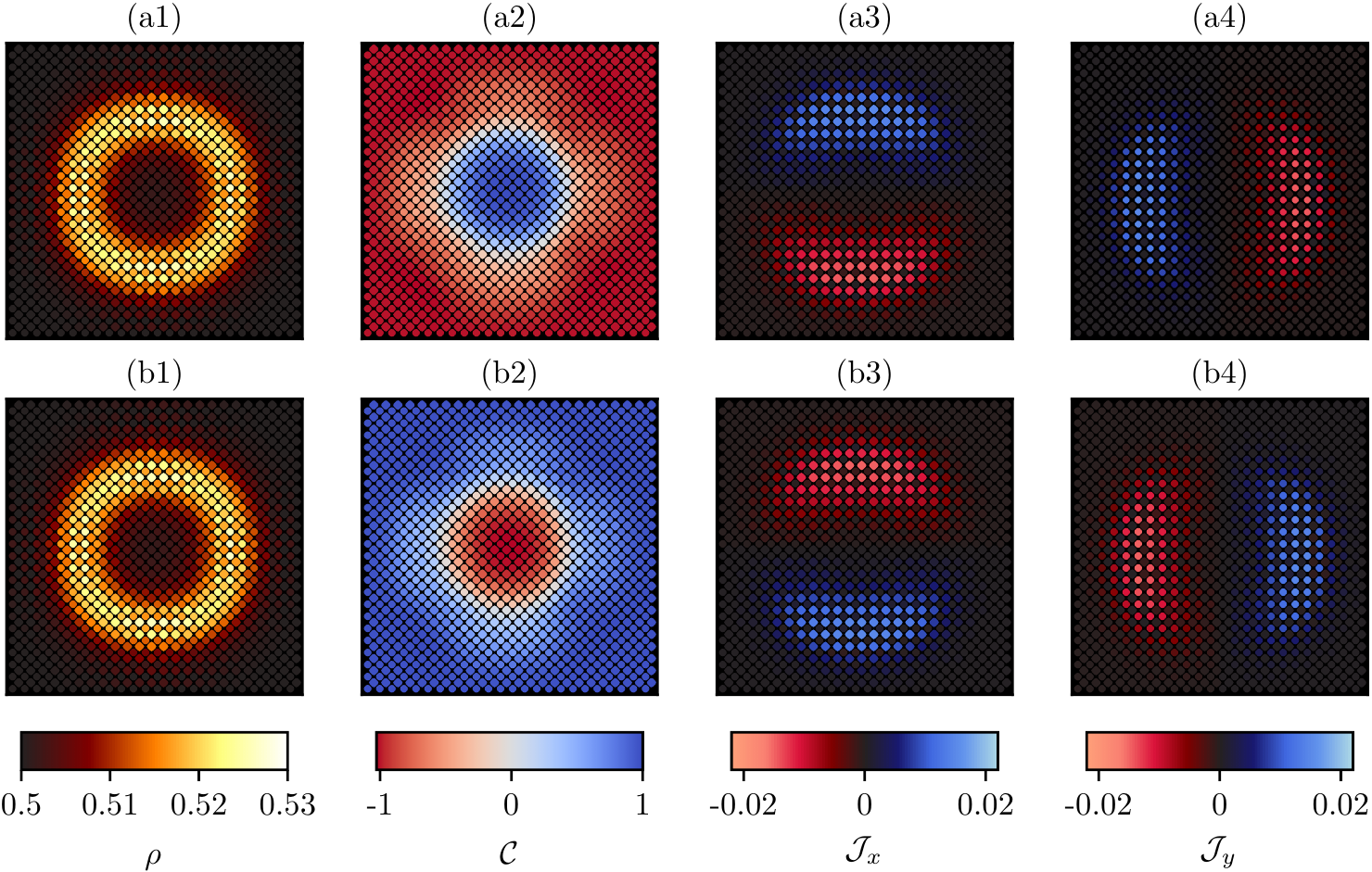}
\caption{Comparison between two degenerate UHF solutions corresponding to ring domain walls (a) and (b). (a1)-(b1) Density profiles, identical for the two solutions. (a2)-(b2) Local Chern number, which shows opposite SSB (topological) sectors for each solution. (a3)-(b3) Current in the $x$ direction. (a4)-(b4) Current in the $y$ direction.  The currents reveal that the chirality of the ring edge states on each solution is reversed with respect to the other, due to the change in the values of the inner (outer) local Chern number. }
\label{fig:ring_sectors}
\end{figure}

As explained in the main text, the spontaneous TRSB in the quantum anomalous Hall phase has two symmetry sectors at half-filling, which lead to two ground states with opposite Chern numbers. In the case of inhomogeneous topological phases, such as the ring domain solution presented in the main text, a first thing to notice is that they are highly degenerate as they spontaneously break the translational symmetry of the lattice. Moreover, we can also find a degenerate solution in which the local TRSB sectors are reversed. Figure~\ref{fig:ring_sectors} shows the comparison between these two degenerate ring solutions that are connected by an inversion of the local TRSB sectors. As can be seen in Figs.~\ref{fig:ring_sectors}(a1)-(b1), the change in the TRSB sectors does not affect the density profile of the solution. However, we can observe in Figs.~\ref{fig:ring_sectors}(a2)-(b2) that the values of the local Chern number $\mathcal{C}$ are reversed in the inner and outer regions of the ring. The latter leads to opposite chiralities of the edge states inside the ring [see Figs.~\ref{fig:ring_sectors}(a3)-(b3)-(a4)-(b4)].

\section{Expression of the current operator}\label{app:current}

To compute the current operator we use the continuity equation for the fermion density at each lattice site:

\begin{equation}
\frac{\text{d}}{\text{dt}}\mean{\cdop_i\cop_i}=\im \mean{\left[H,\cdop_i\cop_i\right]}.
\end{equation}
Now, the non-diagonal terms of $H$ can be written as 
\begin{equation}
\sum_{m,n}(t_{m,n}\cdop_m\cop_n+\text{h.c.}),
\end{equation}
leading to 

\begin{equation}
\frac{\text{d}}{\text{dt}}\mean{\cdop_i\cop_i}=\im\sum_{m,n} \left(t_{m,n}\mean{\left[\cdop_m\cop_n,\cdop_i\cop_i\right]}-\text{h.c.}\right)=-2\ \Im\left({t_{m,n}\mean{\left[\cdop_m\cop_n,\cdop_i\cop_i\right]}}\right).
\end{equation}
Expanding the commutator in terms of anticommutators we easily find:

\begin{equation}
t_{m,n}\left[\cdop_m\cop_n,\cdop_i\cop_i\right]=t_{m,n}\left(\cdop_m\cop_i\{\cop_n,\cdop_i\}-\cdop_i\cop_n\{\cdop_m,\cop_i\}\right)=t_{m,n}\left(\cdop_m\cop_i\delta_{n,i}-\cdop_i\cop_n\delta_{m,i}\right)=t_{i+b,i}\cdop_{i+b}\cop_i-t_{i,i-b}\cdop_i\cop_{i-b},
\end{equation}
where $b$ gives the bond direction, which can be $x+y$ and $x-y$ for nearest-neighbors, and $2x$ or $2y$ for next-to-nearest-neighbors. Introducing this into the continuity equation we get to:

\begin{equation}
\frac{\text{d}}{\text{dt}}\mean{\cdop_i\cop_i}=-2\ \Im \mean{t_{i+b,i}\cdop_{i+b}\cop_i-t_{i,i-b}\cdop_i\cop_{i-b}}.
\end{equation}
At this point, we identify the outgoing current in the $b$ direction as $\mathcal{J}^{b\rightarrow}_{i}\equiv 2\ \Im \mean{t_{i+b,i}\cdop_{i+b}\cop_i}$ and the ingoing as $\mathcal{J}^{b\leftarrow}_{i}\equiv 2\ \Im \mean{t_{i,i-b}\cdop_{i}\cop_{i-b}}$, so that we can write the continuity equation simply as

\begin{equation}
\frac{\text{d}}{\text{dt}}\mean{\cdop_i\cop_i}=\sum_b  \mathcal{J}^{b\leftarrow}_{i} - \sum_b  \mathcal{J}^{b\rightarrow}_{i}.
\end{equation}
Now, we notice that in our model the Hartree-Fock correction to $t_{m,n}$ does not affect the current, and therefore we should use the bare hopping elements $t_{m,n}^0$,

\begin{equation}
\mathcal{J}^{b\rightarrow}_{i} = 2\ \Im \left[\left(t_{i+b,i}^0-V_b\mean{\cop_i\cdop_{i+b}}\right)\mean{\cdop_{i+b}\cop_i}\right]=2\ \Im \left(t_{i+b,i}^0\mean{\cdop_{i+b}\cop_i}-V_b|\mean{\cdop_{i+b}\cop_i}|^2\right)=2\ \Im \ t_{i+b,i}^0\mean{\cdop_{i+b}\cop_i}.
\end{equation}
In the topological homogeneous phase, we have closed loops with non-zero $\mathcal{J}_{ij}$ for nearest-neighbors sites, and $\mathcal{J}_{ij}=0$ for next-to-nearest-neighbors (sites belonging to the same sublattice). In the case of domain walls separating two regions with opposite Chern number, we expect the appearance of chiral currents. We will therefore look for non-zero values of $\mathcal{J}_{ij}^{A(B)}$ in these areas, to identify such currents.

\section{Configuration-interaction}\label{sec:configuration_interaction}
Here we use the Configuration Interaction (CI) method \cite{SlaterOverlap, Guinea99, PhysRevB.83.165118,PhysRevB.89.155141}  to analyze the stability of the UHF solutions presented in the main text. In the CI method, one first obtains a set $\{\Phi_j(N)\}$ of UHF solutions with low energy and a fixed number of particles $N$. In the many-body basis, each of the the solutions is given by the Slater determinant (SD) of the occupied single-particle UHF states. Then, one uses the variational ansatz 
\begin{equation}
\Psi(N)=\sum_j \alpha_j \Phi_j(N),
\end{equation}
to minimize the energy of the system. Notice that the many-body wave-functions $\{\Phi_j(N)\}$ are not orthogonal. The overlap matrix between two SDs is given by \cite{SlaterOverlap}
\begin{equation}
S_{ij}\equiv \braket{\Phi_i|\Phi_j} = 
\begin{vmatrix}
\braket{\phi^i_1|\phi^j_1} & \dots & \braket{\phi^i_1|\phi^j_N} \\
\vdots & \ddots & \vdots \\
\braket{\phi^i_N|\phi^j_1} & \dots & \braket{\phi^i_N|\phi^j_N}
\end{vmatrix}.
\end{equation}
The biggest numerical cost of the CI method is to compute the matrix elements $H_{ij}=\bra{\Phi_i}H\ket{\Phi_j}$ of the Hamiltonian \eqref{eq:hamiltoniansup}. We first split it in the kinetic energy term $T$ and the interaction term $C$:

\begin{equation}
H = T+C \ .
\end{equation}
The matrix elements corresponding to $T$, which are quadratic in fermionic operators, are then given by

\begin{equation}
T_{ij} = 
\begin{vmatrix}
\bra{\phi^i_1}T\ket{\phi^j_1} & \dots & \braket{\phi^i_1|\phi^j_N} \\
\vdots & \ddots & \vdots \\
\bra{\phi^i_N}T\ket{\phi^j_1} & \dots & \braket{\phi^i_N|\phi^j_N}
\end{vmatrix}+ \dots + 
\begin{vmatrix}
\braket{\phi^i_1|\phi^j_1} & \dots & \bra{\phi^i_1}T\ket{\phi^j_N} \\
\vdots & \ddots & \vdots \\
\braket{\phi^i_N\phi^j_1} & \dots & \bra{\phi^i_N}T\ket{\phi^j_N}
\end{vmatrix}.
\end{equation}
On the other hand, $C$ contains two-body terms of the form $n_kn_l$, whose matrix elements read

\begin{equation}
\bra{\Phi_i}n_kn_l\ket{\Phi_j} = 
\begin{vmatrix}
\bra{\phi^i_1}n_k\ket{\phi^j_1} & \bra{\phi^i_1}n_l\ket{\phi^j_2} & \dots & \braket{\phi^i_1|\phi^j_N} \\
\vdots & \ddots &\dots & \vdots \\
\bra{\phi^i_N}n_k\ket{\phi^j_1} & \bra{\phi^i_N}n_l\ket{\phi^j_1}&\dots &  \braket{\phi^i_N|\phi^j_N}
\end{vmatrix}+ \text{permutations}.
\end{equation}
Once $H$ is computed, it has to be diagonalized, taking into account that we are working in a non-orthogonal basis. That is, we want to find the energies $\varepsilon$ and eigenvectors $\vec{\alpha}$ satisfying 

\begin{equation}
(H-\varepsilon S)\vec{\alpha}=0
\end{equation}

To find the energies of the system in this non-orthogonal basis we proceed as follows:

\begin{enumerate}

\item Diagonalize $S$, as $\sigma =D^\dagger S D$. Notice that $\sigma \geq 0$ due to the fact that $S$ is an overlap matrix. A zero eigenvalue indicates that one of the $\Phi_j$ is a linear combination of the others, and should be removed from the set. 
\item Construct $A_{ij}=D_{ij}/\sqrt{\sigma}$. This means that $A^\dagger S A=I$. 
\item By defining $A\vec{\alpha}\equiv\vec{c}$, and $\tilde{H}\equiv A^\dagger H A$, the eigenvalue problem has the standard form $(\tilde{H}-\varepsilon)\vec{c}=0$.

\end{enumerate}

\subsection{Polaron band in the QAH phase}

The CI method can be used to restore the translational invariance of the system, which is spontaneously broken in the unrestricted Hartree-Fock method. To do so, we take one Hartree-Fock solution and perform all possible unit-cell translations. Notice that the SDs obtained through this procedure have the same energy. Then, one finds the band structure of the single-particle excitation $E(\mathbf{k})$ by diagonalizing $H$ in this non-orthogonal basis of translated states.
\begin{figure}[h!]
\centering
\vspace{-1mm}
\includegraphics[width=
0.255\textwidth]{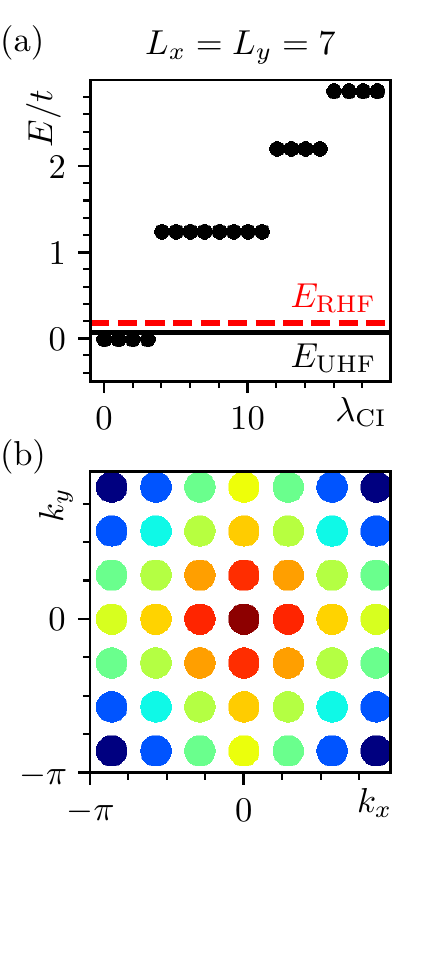}\hspace{-2mm}
\includegraphics[width=
0.255\textwidth]{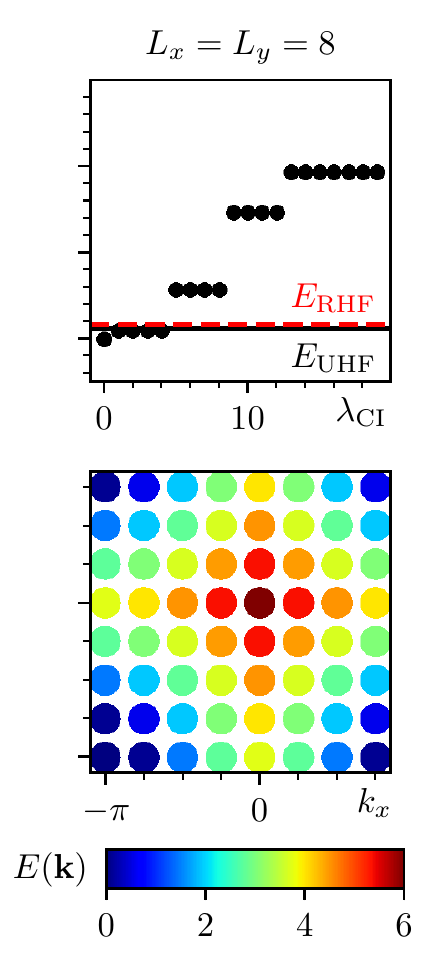}
\includegraphics[width=
0.255\textwidth]{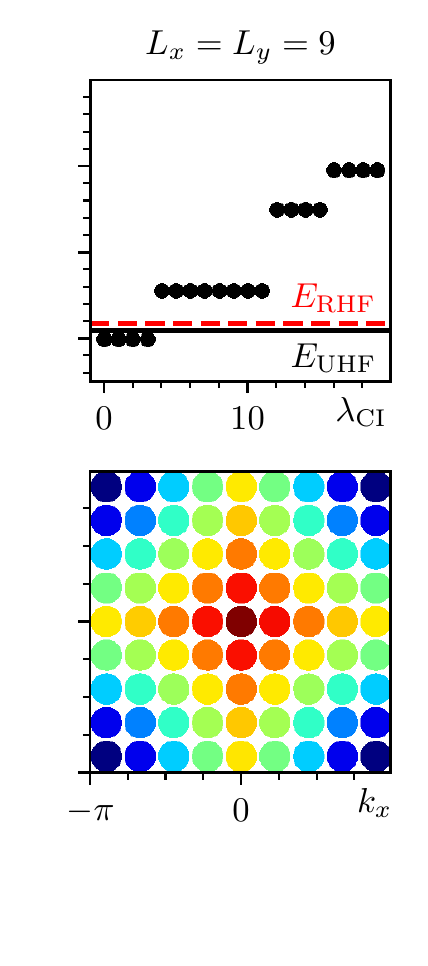}
\caption{Results of the CI method applied to the extended polaron obtained in the UHF ansatz ($\delta=1$), and with the same hopping and interaction parameters as in the main text. (a) CI spectrum for three different lattice sizes. The unrestricted (restricted) Hartree-Fock energies are indicated as solid black (dashed red) lines. For each lattice size, the zero of energy is set as the CI ground state energy. (b)  CI band structure.}
\label{fig:band_polaron}
\end{figure}

\section{Quantum simulation with Rydberg atoms in an optical lattice }\label{sec:Rydberg-atoms}

Here, we provide some details on the envisioned implementation scheme using a setup of Rydberg-dressed cold atoms in an optical lattice. For the realisation of the interacting Hamiltonian of Eq.~(2) it is essential to be able to (i) induce the repulsive interatomic interactions of strength $V_1$ between particles on nearest-neighbor and $V_2$ between next-to-nearest neighbor sites of the lattice. Furthermore, (ii) these effective interaction couplings need to be on the same order of magnitude as the hopping energy scale $t$ of the free Hamiltonian of Eq.~(1), and (iii) the system needs to be prepared at sufficiently low temperatures $T$, below the bulk gap $\Delta$ in order to observe the predicted topological phases and below the energy differences between different solutions $\Delta_{s}$ to detect the spatial structures.

(i) We propose to induce the interaction via off-resonant laser coupling of the fermionic electronic ground state atoms to electronic high-lying Rydberg states. The latter states exhibit bare dipolar interactions that can be orders of magnitude stronger (GHz) than comparable van-der-Waals interactions between ground state atoms at comparable distances ($\mu m$) \cite{saffman-rmp-82-2313}. The idea is to use weak, far off-resonant laser coupling between ground and Rydberg states to admix a small, controllable part of Rydberg state character to the electronic ground states. Thereby, the dressed ground state atoms, inherit the Rydberg interactions, however at a strongly reduced strength, which is compatible with kinetic energies of the atoms hopping in the lattice.

The internal dynamics of the laser-driven interacting atoms is described by the Hamiltonian $H = \sum_i H_i + \sum_{i<j} H_{ij}$, where the single-particle terms $H_i = (\Omega \ket{r}\bra{g}_i + h.c.) + \Delta \ket{r} \bra{r}_i$ account for the laser coupling (in dipole and rotating wave approximation) of electronic ground state atoms in $\ket{g}$ to a Rydberg state $\ket{r}$. Here, $\Omega$ and $\Delta$ denote the laser Rabi frequency and detuning, respectively. The two-body interactions between atoms excited to a Rydberg state $\ket{r}$ are described by $H_{ij} = V_{ij} \ket{r}\bra{r}_i \otimes  \ket{r}\bra{r}_j$. Here, we focus on Rydberg states with zero orbital angular momentum $l=0$ ($s$-states) \cite{gallagher-book} with repulsive and spatially isotropic van-der-Waals interactions $V_{ij} > 0$, $V_{ij} = C_6 |\mathbf{r}_i - \mathbf{r}_j|^{-6}$, which quickly fall off with the inter-particle distance. 

For far off-resonant laser dressing, $\Omega/\Delta \ll 1$, the Rydberg states can be adiabatically eliminated so that (in fourth-order perturbation theory in $\Omega/\Delta$) a pair of dressed ground state atoms undergoes a distance-independent interaction energy shift (see e.g. Ref.~\cite{PhysRevA.86.053618} for details) $(\Delta E)_{\ket{gg}}$
\begin{equation}
	\label{eq:real_part}
	(\Delta E)_{\ket{gg}} = 2 \frac{\Omega^4}{\Delta^3} \left[ 1+ \frac{2\Delta}{V_{ij}} \right]^{-1}.
\end{equation}
\begin{figure}[h!]
	\centering
	\includegraphics[width=
	0.2\textwidth]{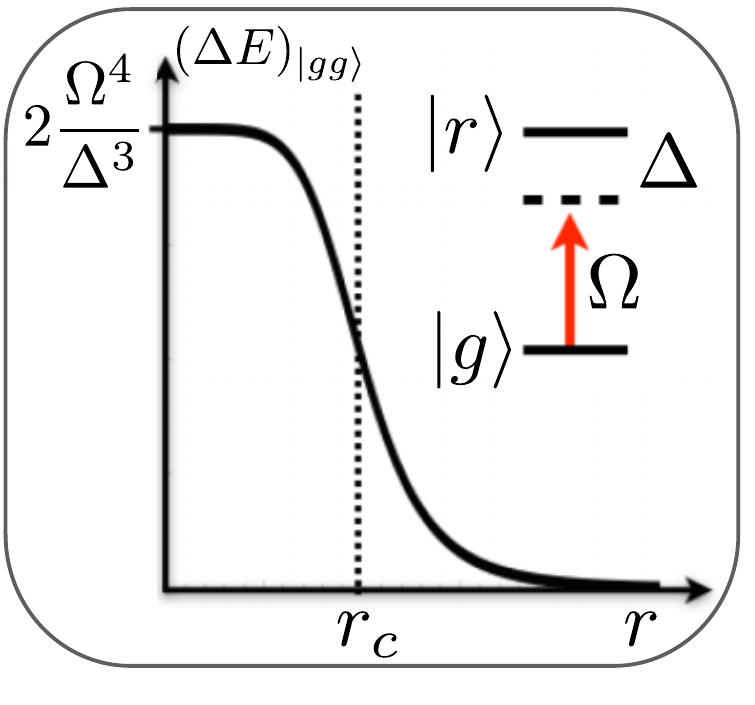}
	\caption{Schematic of the effective interatomic interaction potential between laser-dressed ground state atoms.}
	\label{fig:Rydberg-dressing}
\end{figure}
This effective laser-induced Born-Oppenheimer potential is shown in Fig.~\ref{fig:Rydberg-dressing} and exhibits a characteristic plateau-like potential structure \cite{Nath-2010,Pupillo-2010}. Importantly, at small distances, below a critical inter-atomic distance $r_c = (C_6 /2 \Delta)^{1/6}$, which is given by the condition $2 \Delta \approx V_{ij}$, atoms interact strongly and the effective energy shift $(\Delta E)_{\ket{gg}} = 2 \, \Omega^4/\Delta^3$ becomes distance-independent. On the other hand, for $r \gg r_c$, interactions are weaker and the effective interaction profile falls off rapidly, $(\Delta E)_{\ket{gg}}  = (\Omega/\Delta)^4 V_{ij}$, according to the characteristic $1/r^6$ van-der-Waals distance dependence, however, also strongly suppressed by the pre-factor $(\Omega/\Delta)^4$. The idea is to choose parameters such that the NN and NNN inter-particle distances lie within the critical radius $r_c$, so that the corresponding effective interaction strengths $V_1$ and $V_2$ are of comparable magnitude, whereas the effect longer-range interactions ($V_i$, $i \geq 3$) take place at distances greater than $r_c$, where interactions fall off rapidly. 

The optimal choice of parameters depends on the specific experimental setup  and fermionic atomic species used. However, reaching realistic conditions is certainly feasible: the critical length scale (Rydberg blockade radius) $r_c$ needs to be slightly larger than the lattice spacing $a$ ($r_c \gtrsim a$), typically $a \approx$ 500\,nm. For Rydberg $s$ states of principal quantum number $n$ around $n=30$, with van-der-Waals interactions of about $2 \pi \hbar \times$ 100 MHz at nearest-neighbor distance, one can choose a detuning of $\Delta = 2 \pi \hbar \times$ 50 MHz, ten times larger than the laser Rabi frequency $\Omega = 2 \pi \hbar \times$ 5 MHz. This set of parameters results in effective interaction shifts between dressed ground state atoms of $(\Delta E)_{\ket{gg}} \approx 2\pi \times 10 \,\text{kHz}$, which is compatible with tunneling rates $t$ of a few kHz, fulfilling the second requirement (ii) as discussed above.

The finite lifetime of the Rydberg states of typically a few tens of microseconds for $n=30$ Rydberg $s$-states at room temperature \cite{gallagher-book,saffman-rmp-82-2313} induces unwanted effective decoherence for the dressed ground state atoms. However, the effective interaction energy scale (or associated frequency) is much larger than the effective decay rate $\gamma_{\rm eff} = (\Omega/\Delta)^2 \gamma_r = 2 \pi \times 50$\,Hz. 

Regarding the requirement (iii), the parameters chosen in the main text yield a bulk gap of $\Delta \simeq t$ (Fig. 2(a) of the main text). Furthermore, the energy difference between the spatially inhomogeneous solutions and the corresponding homogeneous ones is $\Delta_{\text{DW}} \simeq 0.6t$ for the ring-shaped domain wall and $\Delta_{\text{pol}}\simeq 0.1t$ for individual polarons.  For typical values of the hopping $t \simeq 1\ \text{kHz}$, these yield the critical temperatures  $\Delta/K_B \sim 50$ nK, $\Delta_{\text{DW}}/K_B \sim 5$ nK, $\Delta_{\text{pol}}/K_B \sim 1$ nK.

\bibliographystyle{apsrev4-1}